\documentclass[aps,twocolumn,prl,superscriptaddress,amsmath,showpacs,tightenlines]{revtex4}
\usepackage{epsfig,graphicx,times}
\usepackage{amstext}
\usepackage{amsmath}
\usepackage{amssymb}
\usepackage{graphicx}
\usepackage{latexsym}
\usepackage{bm}

\begin{document}
\title{Optical selection rules and phase-dependent adiabatic state control in a superconducting quantum circuit}
\author{Yu-xi Liu}
\affiliation{Frontier Research System, The Institute of Physical
and Chemical Research (RIKEN), Wako-shi 351-0198, Japan}
\author{J. Q. You}
\affiliation{Frontier Research System, The Institute of Physical
and Chemical Research (RIKEN), Wako-shi 351-0198, Japan}
\affiliation{Department of Physics and Surface Physics Laboratory
(National Key Laboratory), Fudan University, Shanghai 200433,
China}
\author{L. F. Wei}
\affiliation{Frontier Research System, The Institute of Physical
and Chemical Research (RIKEN), Wako-shi 351-0198, Japan}
\affiliation{IQOQI, Department of Physics, Shanghai Jiaotong
University, Shanghai 200030, China }
\author{C. P. Sun}
\affiliation{Frontier Research System, The Institute of Physical
and Chemical Research (RIKEN), Wako-shi 351-0198, Japan}
\affiliation{Institute of Theoretical Physics, Chinese Academy of
Sciences, Beijing, 100080, China}
\author{Franco Nori}
\affiliation{Frontier Research System, The Institute of Physical
and Chemical Research (RIKEN), Wako-shi 351-0198, Japan}
\affiliation{MCTP, Physics Department, CSCS, The University of
Michigan, Ann Arbor, Michigan 48109, USA}
\date{\today}

\begin{abstract}
We analyze the optical selection rules of the microwave-assisted
transitions in a flux qubit superconducting quantum circuit (SQC).
We show that the parities of the states relevant to the
superconducting phase in the SQC are well-defined when the
external magnetic flux $\Phi_{\rm e}=\Phi_{0}/2$, then the
selection rules are same as the ones for the electric-dipole
transitions in usual atoms. When $\Phi_{\rm e}\neq \Phi_{0}/2$,
the symmetry of the potential of the artificial ``atom'' is
broken, a so-called  $\Delta$-type ``cyclic" three-level atom is
formed, where  one- and two-photon processes can coexist. We study
how the population of these three states can be selectively
transferred by adiabatically controlling the electromagnetic field
pulses. Different from $\Lambda$-type atoms, the adiabatic
population transfer in our three-level $\Delta$-atom can be
controlled not only by the amplitudes but also by the phases of
the pulses.

\end{abstract}


\maketitle \pagenumbering{arabic}

{\it Introduction.---} Analogous to natural atoms, superconducting
quantum circuits (SQCs) can possess discrete levels. Such
artificial atoms provide a promising ``hardware" for quantum
information processing. Micro-chip electric
circuits~\cite{you,wallraff} show that quantum optical effects can
also appear in artificial atoms, allowing quantum information
processing in such circuits by using cavity quantum
electrodynamics (e.g.,~\cite{you,wallraff,girvin,yang,amz}).

Quantum optical technology, developed for atomic systems, can be
used to manipulate quantum states of artificial atoms. For
example, the selective population transfer~\cite{jpm} based on the
stimulated Raman adiabatic passage (STIRAP) with $\Lambda$-type
atoms~\cite{scullybook,zoller} has been applied to superconducting
flux qubits~\cite{yang,zhou}. How to probe the decoherence of flux
qubits by using electromagnetically induced transparency has also
been investigated in a SQC~\cite{orlando3} formed by a loop, with
three Josephson junctions.

We investigate a generalized STIRAP approach for the novel type of
artificial atoms presented here. It is well known that the
parities of eigenstates are well-defined for usual atomic systems.
Due to their atomic symmetry, described by $SO(3)$ or $SO(4)$,
one-photon transitions between two energy levels require that the
two corresponding eigenstates have opposite parities; but a
two-photon process needs these states to have the same parities.
However, this situation can be significantly changed for
artificial atoms due to its easily controllable (by the external
magnetic flux $\Phi_{\rm e}$) effective potential.

Here, we focus on the flux qubit circuit~\cite{orlando,you2},
analyzing its parity. When $\Phi_{\rm e}=\Phi_{0}/2$, the qubit
potential energy of the superconducting phases is symmetric, and
the interaction Hamiltonian between the time-dependent microwave
field and the qubit also has a well-defined parity. In this case,
{\it the optical selection rules} of the microwave-assisted
transitions between different qubit states are {\it the same as
for the electric-dipole ones} in usual atoms: {\it one- and
two-photon transitions cannot coexist}. But if $\Phi_{\rm
e}\neq\Phi_{0}/2$, the symmetries of  both the potential and the
interaction Hamiltonian are broken. Then the selection rules do
not hold, and an unusual phenomenon appears: {\it one- and
two-photon processes can coexist}~\cite{shapiro1,shapiro}. In this
case, all transitions between any two states are possible. Then
{\it the population can be cyclically transferred} with the
assistance of (the amplitudes and/or phases of) microwave pulses.
Thus, we achieve a {\it pulse-phase-sensitive adiabatic
manipulation of quantum states} in this three-level artificial
atom. Usually, only the amplitude was considered for adiabatic
control.

{\it Broken symmetry of the superconducting phase and selection
rules.---} We consider a qubit circuit composed of a
superconducting loop with three Josephson junctions
(e.g.,~\cite{orlando} and ~\cite{orlando2}). The two larger
junctions have equal Josephson energies $E_{\rm{J1}
}=E_{\rm{J2}}=E_{\rm{J}}$ and capacitances
$C_{\rm{J1}}=C_{\rm{J2}}=C_{\rm{J}}$, while for the third
junction: $E_{\rm J3}=\alpha E_{\rm J}$ and $C_{\rm J3}=\alpha
C_{\rm J}$, with $\alpha <1$. The Hamiltonian is

\begin{equation*}
H_{0}\,\,=\,\,\frac{P_{p}^{2}}{2M_{p}}+\frac{P_{m}^{2}}{2M_{m}}+U(\varphi
_{p},\varphi _{m}),
\end{equation*}
with $M_{p}=2C_{\rm J}(\Phi_{0}/2\pi)^2$ and
$M_{m}=M_{p}(1+2\alpha)$. The effective potential $U(\varphi
_{p},\varphi _{m})$ is
%\begin{eqnarray}
$U(\varphi _{p},\varphi _{m})\,\,=\,\,2E_{\rm{J}}(1-\cos \varphi
_{p}\cos \varphi _{m})+\alpha E_{\rm J}[1-\cos (2\pi f+2\varphi
_{m})]$ where $\varphi _{p}=(\varphi _{1}+\varphi _{2})/2$ \, and
\,$ \varphi _{m}=(\varphi _{1}-\varphi _{2})/2$ are defined by the
phase drops $\varphi _{1}$ and $\varphi _{2}$ across the two
larger junctions; $f=\Phi _{\rm e}/\Phi _{0}$ is the reduced
magnetic flux.

\begin{figure}
\includegraphics[bb=30 229 578 755, width=8.8 cm]{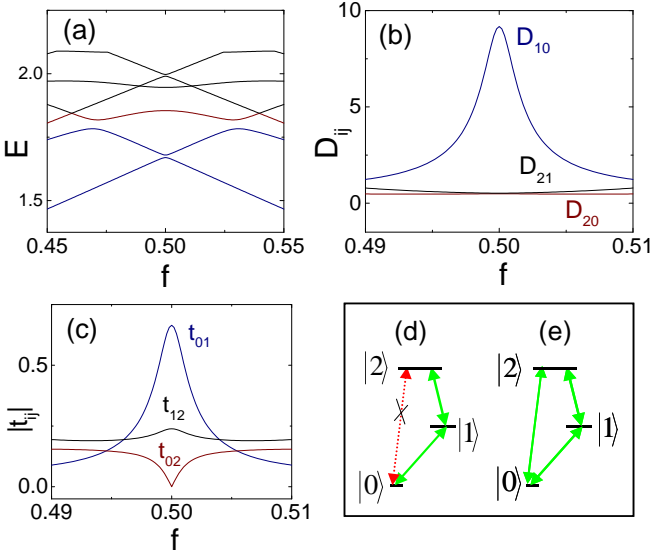}
\vspace{-0.5cm} \caption{(Color online). (a) Energy levels, in
units of $E_{\rm{J}}$, of the flux qubit vs reduced flux $f$ for
states $|0\rangle$ to $|5\rangle$. (b) $f$-dependent ratios of
transition frequencies:
$D_{ij}=(\varepsilon_{3}-\varepsilon_{2})/(
\varepsilon_{i}-\varepsilon_{j})$ with $j < i < 3$. (c) Moduli
$|t_{ij}|$ of the transition matrix elements between states
$|i\rangle$ and $|j\rangle$ vs $f$,  for the lowest three levels.
Transition diagram of three energy levels vs $f$. (d) corresponds
to $f=0.5$ and (e) to $f\neq0.5$, where the green (red) line means
allowed (forbidden) transitions. (e) A triangle-shaped or
$\Delta$-type energy diagram for $f\neq 0.5$, where all photon
assisted transitions are possible and the electric dipole
selection rules do not hold.}\label{fig1}
\end{figure}

Figure~\ref{fig1}(a) summarizes numerical results of the
$f$-dependent spectrum for $H_{0}$ up to the sixth eigenvalue,
with $\alpha =0.8$ and $E_{\rm J}=40E_{\rm c}$ (as
Ref.~\cite{orlando}). Fig.~\ref{fig1}(a) shows that near the point
$f=0.5$, e.g., $f=0.496$, the lowest three energy levels are well
separated from other higher energy levels. Then the lowest two
energy levels form a two-level artificial atom, called qubit, with
an auxiliary third energy level. Fig.~\ref{fig1}(b) plots
$f$-dependent ratios
$D_{ij}=(\varepsilon_{3}-\varepsilon_{2})/(\varepsilon_{i}-\varepsilon_{j})$
of the transition frequency between the fourth and third energy
levels with the other three among the lowest three energy levels
($j<i<3$). It is found that $D_{ij}\neq 1$, so the transition
frequencies of different eigenstates are not equal when f is near
0.5. Then when we manipulate the lowest three states, the fourth
state will be well separated and not be populated.

When a time-dependent microwave electromagnetic field $\Phi _{\rm
a}(t)$ is applied through the loop, photon-assisted transitions
occur. For small $\Phi _{\rm a}(t)$, the $\varphi _{m}-$dependent
perturbation Hamiltonian reads
%\begin{subequations}
%\begin{equation}
$H_{1}(\varphi_{m}, t)\,=\,I\Phi _{\rm a}(t)\,=\,I\Phi _{\rm
a}^{(0)}\cos (\omega _{ij}t),$ %\label{eq:2}
%\end{equation}
where
%\begin{equation}
$I\,=\,-(2\pi \alpha E_{\rm J}/\Phi _{0})\sin (2\pi f+2\varphi
_{m})$
%\end{equation}
%\end{subequations}
is the circulating supercurrent when $\Phi _{\rm a}(t)=0$, and the
amplitude $\Phi _{\rm a}^{(0)}$ is now assumed to be
time-independent. The transitions are determined by the matrix
elements $t_{ij}=\langle i|I\Phi _{\rm a}(t)|j\rangle $ for the
$i$th and $j$th eigenstates $|i\rangle $ and $|j\rangle $, with
eigenvalues $\varepsilon _{i}$ and $\varepsilon _{j}$,
respectively.

Analytically, the potential $U(\varphi _{p},\varphi _{m})$ is an
even function of $\varphi _{p}$ and $\varphi_{m}$ when $2f$ is an
integer; however $H_{1}(\varphi _{m},t)$ is an odd function of
$\varphi_{m}$. Thus, at these specific points, the {\it parities}
of $H_{1}(\varphi _{m},t)$ and the eigenstates of $H_{0}$ are
well-defined. Then, the selection rules for the transition matrix
elements $t_{ij}$ at these points have the same behavior as the
ones for the {\it electric-dipole transitions} in atoms due to
{\it the odd parity} of $H_{1}(\varphi _{m},t)$. However, if $f$
deviates from these points, the symmetries are broken and thus the
dipole selection rules do not hold. Fig.~\ref{fig1}(c) shows the
$f$-dependent transition elements $|t_{ij}|$ for resonant
microwave frequencies between any two of the lowest three states:
$|0\rangle $, $|1\rangle $ and $|2\rangle $. It shows that the
transition $|0\rangle \leftrightarrow |2\rangle $ is forbidden at
$f=0.5$, but the transitions $|1\rangle \leftrightarrow |2\rangle
$ and $|0\rangle \leftrightarrow |1\rangle $ reach their maxima at
this point. Then these three states have ladder-type or $\Xi$-type
transitions~\cite{jpm,scullybook}, as shown in Fig.~\ref{fig1}(d).
The states $|0\rangle $ and $|2\rangle $ have the same parities
when $f=0.5$. Fig~\ref{fig1}(c) also demonstrates that {\it all}
photon-assisted transitions are possible when $f\neq 0.5$, as in
Fig.~\ref{fig1}(e), showing what we call $\Delta $-type (or
triangle-shaped) transitions. So a $\Lambda$-type ``atom",
allowing transitions $|0\rangle \leftrightarrow |2\rangle$ and
$|1\rangle \leftrightarrow |2\rangle$ but prohibiting $|0\rangle
\leftrightarrow |1\rangle$, cannot be realized in this circuit.

{\it Adiabatic energy levels of $\Delta-$artificial atoms.---} In
ladder-type transitions~\cite{jpm,scullybook}, the population of
the lowest state $|0\rangle $ can be adiabatically transferred to
the highest state $|2\rangle $ by applying two appropriate
classical pulses. $\Lambda $-type transitions are usually
required~\cite{jpm,scullybook,zoller} to adiabatically manipulate
the populations of two states $|0\rangle $ and $|1\rangle $ by a
third (auxiliary) state $|2\rangle$.  In contrast with the usual
$\Lambda$-type model~\cite{orlando3}, in our case, when $f\neq
0.5$, the transitions among the lowest three states {\it can be
cyclic}.

Now we consider three electromagnetic pulses  applied through the
loop. We assume $f\neq 0.5$, but near it. The time-dependent flux
is $\Phi _{\rm a}(t)=\sum_{m>n=0}^{2}[\Phi _{mn}(t)e^{-i\omega
_{mn}t}+\Phi _{mn}^{\ast }(t)e^{i\omega _{mn}t}]$, and the $\Phi
_{mn}(t)$ vary slowly on the time-scale of the pulses,  where
$\omega _{mn}$ are the pulse carrier frequencies. If $\omega
_{mn}$ is  resonant or near-resonant with the transitions among
the non-adiabatic (i.e., diabatic) states $|m\rangle $
($m=0,1,2)$, the total Hamiltonian in the interaction picture can
be written under the rotating wave approximation as
%\begin{equation}
$H_{\rm{int}}=\sum_{m>n=0}^{2}\left( \Omega _{mn}(t)e^{i\Delta
_{mn}t}|m\rangle \langle n|+\rm{h.c.}\right),$ %\label{eq:3}
%\end{equation}
where the complex Rabi frequencies $\Omega _{mn}(t)=\langle
m|I\Phi _{mn}(t)|n\rangle $, and the detuning $\Delta _{mn}=\omega
_{m}-\omega _{n}-\omega _{mn}$, with $\omega _{m}=\varepsilon
_{m}/\hbar $.

The instantaneous adiabatic eigenvalues of $H_{\rm int}$ are given
by $E_{k}=2|\Omega(t)|\sqrt{1/3}\cos [(\theta +2(k-1)\pi )/3]$
($k=1,\,2,\,3$). Here,  $\cos \theta =3\sqrt{3} \rm{Re}[\Omega
_{01}(t)\Omega _{12}(t)\Omega _{20}(t)e^{i\omega ^{\prime
}t}]/|\Omega (t)|^{3}$, $|\Omega (t)|^{2}=|\Omega
_{12}(t)|^{2}+|\Omega _{20}(t)|^{2}+|\Omega _{01}(t)|^{2}$,
$\omega ^{\prime }=\Delta _{01}+\Delta _{12}-\Delta _{02}$. The
eigenvalues are sensitive to the total phase $\phi $ of the
product $\Omega _{01}(t)\Omega _{12}(t)\Omega
_{20}(t)=\Omega^{\prime}(t)$ and detunning $\omega ^{\prime }$. It
can be found that $\theta =0$ or $\pi $ when $|\Omega
_{01}(t)|=|\Omega _{12}(t)|=|\Omega _{20}(t)|$ and $\beta =\omega
^{\prime }t+\phi =n\pi $. In such case, there are energy level
crossings and, thus, the adiabatic description of the time
evolution is no longer correct. Comparing with the typical
$\Lambda $-type atom~\cite{jpm}, the time-evolved zero eigenvalue
$E_{3}=0$ (corresponding to $E_{1}=|\Omega (t)|,\,E_{2}=-|\Omega
(t)|$) can also be found for $H_{\rm int}$ when $\beta =(2p+1)\pi
/2$, for integer $p$. In such case, the evolution is adiabatic.

\begin{figure}
\includegraphics[width=8.6 cm]{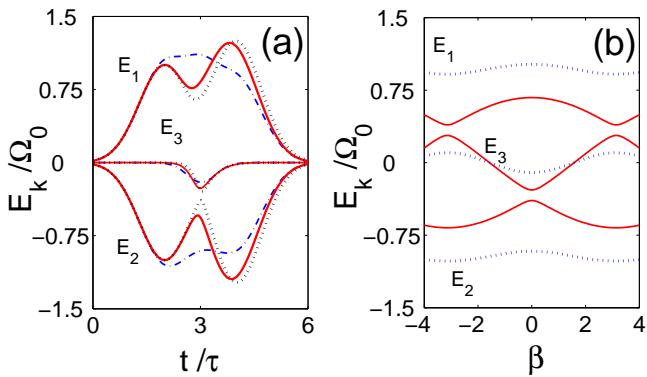}
\caption{(Color online). (a) The rescaled time $t/\tau$ dependence
of the eigenvalues $E_{k}$ of $H_{\rm int}$ with the detuning
$\omega^{\prime}=0$ for different pulse central times:
$\tau_{1}=2\tau_{2}/3=\tau_{3}/2=2 \tau$ (dash-dotted blue
curves); $\tau_{1}=\tau_{3}/2=2
\tau$ and $\tau_{2}=3.7\tau$ (solid red curves), and $\protect\tau%
_{1}=\tau_{2}/2=\tau_{3}/2=2\tau$ (dotted black curves). (b) The
phase-dependent eigenvalues $E_{k}$ at $t=3\protect\tau$ (solid
red curves) and $t=3.5\tau$ (dotted blue curves), with time delays
$\tau_{1}=\tau_{2}/2=\tau_{3}/2=2\tau$. Here eigenvalues are given
in order $E_{1},\, E_{3},\, E_{2}$ from the top curve to the
bottom curve for the same color.} \label{fig2}
\end{figure}

Let us now consider Rabi frequencies $\Omega _{mn}(t)$ as Gaussian
envelops, e.g., $\Omega _{21}(t)=\Omega _{0}\exp
(i\phi_{2}-(t-\tau _{1})^{2}/\tau ^{2})$, $\Omega
_{10}(t)=0.9\,\Omega _{0}\exp [i\phi_{1}-(t-\tau _{2})^{2}/\tau
^{2}]$, and $\Omega _{20}(t)=0.85\,\Omega _{0}\exp
[i\phi_{3}-(t-\tau _{3})^{2}/\tau ^{2}]$, where $\tau$ is the
pulse width, $\phi_{i}$ is the pulse phase, and $\Omega _{0}>0$.
To avoid energy crossings, the pulse central times $\tau _{i}$
($i=1,\,2,\,3$) are chosen such that $|\Omega _{12}|\neq |\Omega
_{01}|\neq |\Omega _{02}|$ during the time evolution. For the case
$\phi =0$ and $\omega ^{\prime }=0$, Fig.~\ref{fig2}(a) shows the
dependence of the eigenvalues $E_{k}$ on the pulse central times
$\tau _{i}$. If $\tau_{i}$'s are equal or nearly equal for two
pulses (e.g. $|\Omega _{01}(t)|$ and $|\Omega _{02}(t)|$), then
two eigenvalues (e.g., $E_{2}$ and $E_{3}$) are closer to each
other when the overlap region between two pulses is large.
Generally, the conditions $\phi =0$ and $\omega ^{\prime }=0$ are
not always satisfied. Combining our expressions for $\phi $ and
$\omega ^{\prime }$, Fig.~\ref{fig2}(b) plots a few snapshots of
the eigenvalues $E_{k}(t)$ versus the phase $\beta $ of
$\Omega^{\prime}(t)$, for the given pulse central times $\tau
_{1}=\tau _{2}/2=\tau _{3}/2=2\tau$. Figure.~\ref{fig2}(b) shows
that two of the eigenvalues $E_{k}$ are closer to each other for
three points ($\phi =-\pi ,\,0,\,\pi$) in the range $|\phi| \leq
\pi$, when $t=3\tau$ and $\omega^{\prime}=0$. If $\omega ^{\prime
}\neq 0$ and $\phi \neq 0$, the phase $\beta=\omega^{\prime}
t+\phi$ always changes in time.  Two eigenvalues are close to each
other when $\beta \cong n\pi $. If the pulses related to
$\Omega_{mn}$ do not have a significant overlap (e.g., when
$t=3.5\tau $), then the three energy levels are well separated. If
the maximum amplitudes of two pulses among $|\Omega _{01}(t)|$,
$|\Omega _{12}(t)|$, and $|\Omega _{02}(t)|$ are the same, then
the central times for these two pulses should be different to
avoid the energy level crossings in the significant overlap area
of these three pluses. However, if two central times, among the
three pluses, are the same, then their maximum amplitudes should
be different in order to guarantee an adiabatic evolution.

{\it Adiabatic control of quantum states.---} The population
transfer from one state to another is implemented via an adiabatic
evolution, where the system remains in the $k$th eigenstate
$|E_{k}\rangle =N_{k}^{-1}(t)(a_{k,2}(t)|2\rangle
+a_{k,1}(t)|1\rangle +|0\rangle )$ of the instantaneous
Hamiltonian $H_{\rm int}$ with
$N_{k}^{2}(t)=|a_{k,2}(t)|^{2}+|a_{k,1}(t)|^{2}+1$. The relative
population amplitudes
%\begin{eqnarray*}
$a_{k,2}(t)=[e^{(i\Delta _{02}t-i\phi _{3})}/b_{k}(t)](|\Omega
_{10}(t)\Omega _{21}(t)|e^{i\beta }+E_{k}|\Omega _{20}(t)|)$ and
$a_{k,1}(t) =[e^{(i\Delta _{12}t-i\phi _{2})}/b_{k}(t)](|\Omega
_{10}(t)\Omega _{20}(t)|e^{-i\beta
}+E_{k}|\Omega _{21}(t)|)$ %\end{eqnarray*}
are determined by the applied pulses. Here
$b_{k}(t)=E_{k}^{2}(t)-|\Omega _{10}(t)|^{2}$.

\begin{figure}
\includegraphics[width=4.3 cm]{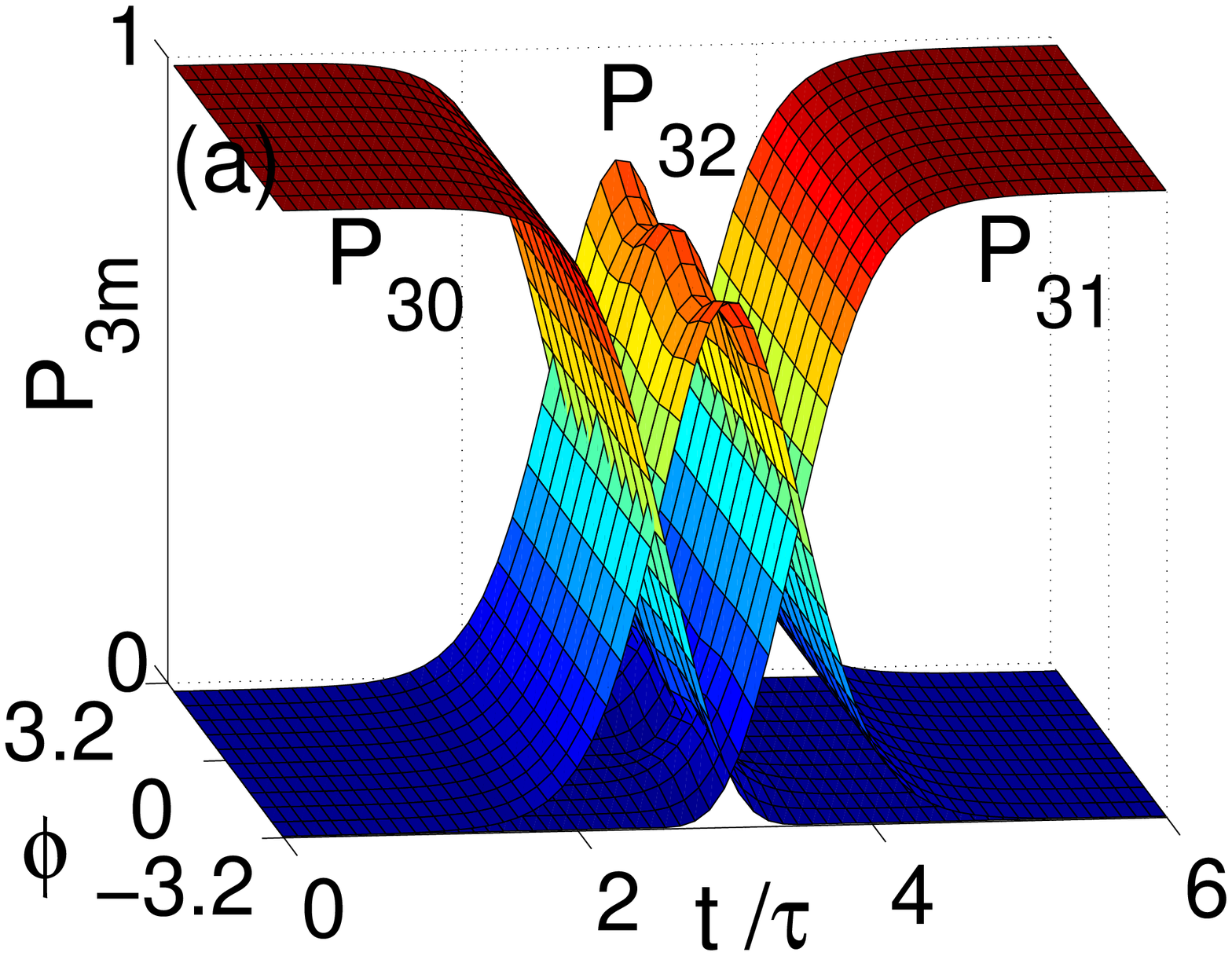}
\includegraphics[width=4.2 cm]{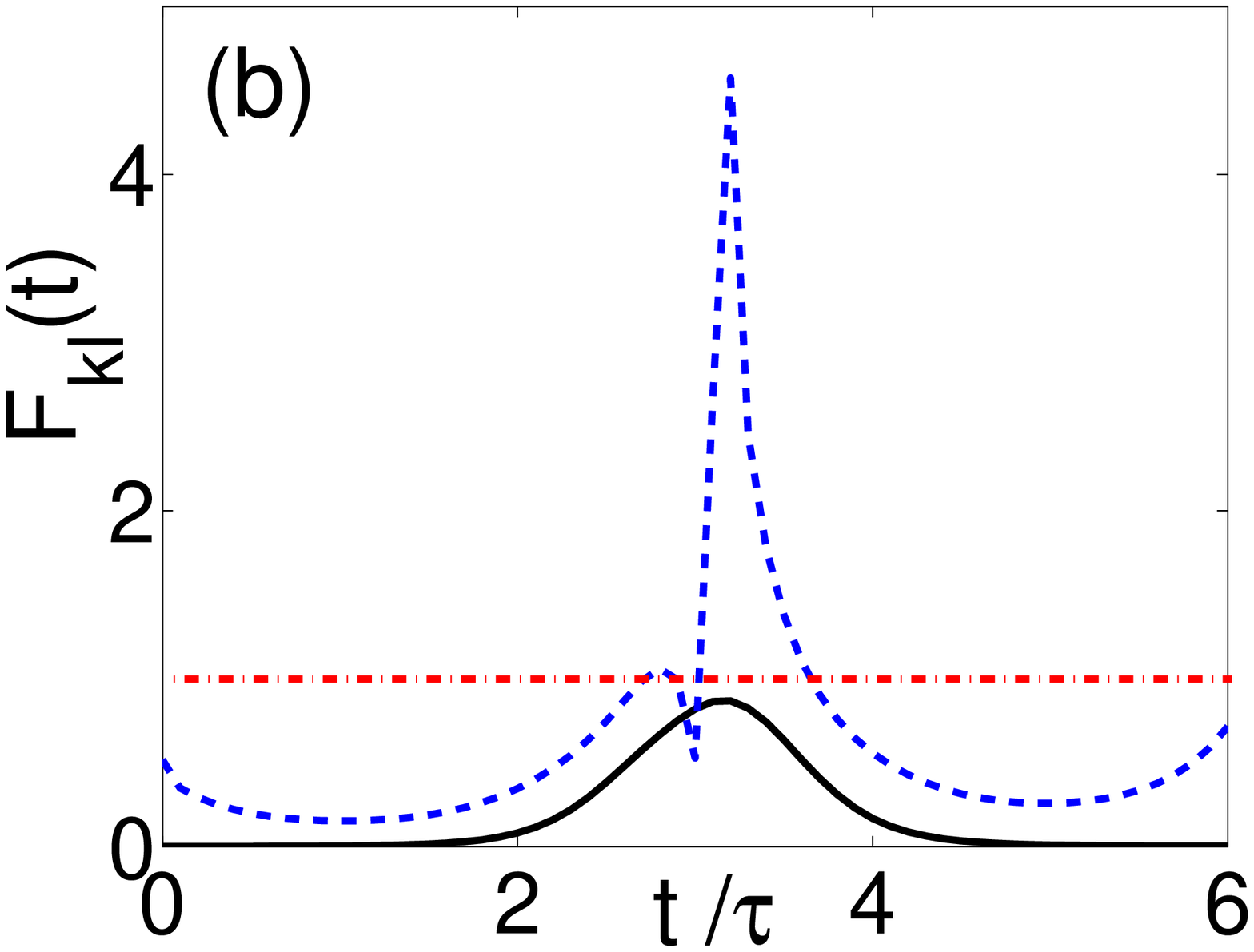}
\caption{(Color online). (a) The rescaled time $t/\protect\tau$
and (phase) $\phi$-dependent probabilities $P_{3m}$ of the
diabatic components in the third adiabatic eigenstate
$|E_{3}\rangle$. (b) The representative top curves of $F_{kl}(t)$
are plotted for two phases $\phi=\pi$ (dashed blue curve) and
$\phi=\pi/2$ (solid black curve) with the pulses given in
Fig.~\ref{fig2} for central times $\tau_{1}=2\tau
_{2}/3=\tau_{3}/2=2\tau$; here, the detunings are taken as
$\Delta_{mn}=0$. The dash-dotted red line denotes $F_{kl}(t)=1$.}
\label{fig3}
\end{figure}

The control of the quantum state can be realized by choosing the
classical pulses such that the diabatic components in the
adiabatic basis states $|E_{k}\rangle $ can be changed during the
adiabatic evolution. For the pulses with the same central times as
in Fig.~(\ref{fig2}), but arbitrary $\omega ^{\prime }$ and $\phi
$, we find that the probabilities $P_{k,m}$ of the diabatic
components $|m\rangle $ (in order $m=0,\,\,1,\,\,2$) for the
eigenstates $|E_{1}\rangle $ and $|E_{2}\rangle $ are:
$(|0\rangle+|1\rangle)/\sqrt{2}$ before the overlap central time
area, but they evolve to $(|1\rangle+|2\rangle)/\sqrt{2}$ after
the overlap central time area. However, it evolves from $
|0\rangle$ to $|1\rangle$ for $|E_{3}\rangle$, which is more
desirable since it coincides with the diabatic (bare) ground state
in the past. As an illustration, Fig.~\ref{fig3}(a) shows the
pulse-phase- and time-dependent probabilities $P_{3,m}(t,\phi)$
for the eigenstate $|E_{3}\rangle$ with $\omega ^{\prime }=0$ and
the pulses given in Fig.~\ref{fig2} with central times $\tau
_{1}=2\tau _{2}/3=\tau _{3}/2=2\tau $. It clearly shows that the
bare ground state $|0\rangle$ coincides with the adiabatic basis
state $|E_{3}\rangle$ when $t\rightarrow 0$. Thus, if the system
evolves adiabatically, the ground state $|0\rangle$ can evolve to
a superposition of three states $|m\rangle $ at $t=3\tau $, then
gradually to a superposition $\alpha |1\rangle +\beta |2\rangle $
of the two upper states, and finally to the first excited state
$|1\rangle$. Generally, the weights of the superpositions
discussed above depend on $\phi$, $\omega ^{\prime }$, and  the
pulse magnitudes $\Omega _{01}(t)$, $\Omega _{12}(t)$, $\Omega
_{20}(t)$. Now let us discuss how the total phase $\phi $ affects
the adiabatic evolution. The adiabatic evolution requires that the
nonadiabatic coupling $\langle E_{k}|\frac{{\rm d}}{{\rm d}
t}{E_{l}}\rangle $ and adiabatic energy differences
$|E_{k}-E_{l}|$ satisfy~\cite{sun-prd} the condition
%\begin{equation}
$F_{kl}= |\langle E_{k}|\frac{{\rm d}}{{\rm d}
t}{E_{l}}\rangle/(E_{k}-E_{l})| \ll 1.$ %\label{eq:5}
%\end{equation}%
%where the dot over $E_{l}$ denotes the time derivative.
For given pulses, it implies that {\it the adiabatic condition}
depends not only on the total detuning $\omega ^{\prime }$ and
individual detunings $\Delta _{mn},$ but also on {\it the total
phase} $\phi$. In Fig.~\ref{fig3}(b), the time evolutions of
$F_{kl}$ are illustrated by showing only one representative top
curve for the central times ($\tau _{1}=2\tau _{2}/3=\tau
_{3}/2=2\tau$) in the resonant case (e.g., $\Delta _{mn}=0$), for
two special phases $\phi =\pi $ and $\phi =\pi /2$.
Fig.~\ref{fig3}(b) shows that the adiabatic condition $F_{kl}\ll
1$ is valid when $\phi =\pi /2$ but it is invalid when $\phi
=\pi$, for fixed envelopes of the Rabi frequencies. By comparing
with Fig.~\ref{fig2}, it is also found that the adiabatic
evolution is invalid even when there are no energy levels crossing
in the overlap area of the pulses, since the non-adiabatic
coupling is very large in this area when $\phi =\pi $.

{\it Discussions and Conclusions.---}We analyzed selection rules
of the artificial atom formed by a SQC, and our analytical results
were numerically confirmed. We find that when the reduced external
magnetic flux $2f$ is an integer, the potential of the artificial
atom has {\it a well-defined symmetry of the superconducting
phases $\varphi_{p}$ and $\varphi_{m}$}, while the interaction
Hamiltonian has {\it odd parity}. Therefore the microwave-assisted
transitions in this artificial atom are the same as
electric-dipole ones. However when $2f$ is not an integer, the
symmetry of the potential is broken and the parity of the
interaction Hamiltonian is also not well defined. In this case,
transitions between any two levels are possible.

Based on the analysis of the selection rules, we discuss the
microwave-assisted adiabatic population transfer among the lowest
three energy levels when $2f$ is not an integer. In this case, the
population of the three levels  can  be transferred cyclically,
and a triangular $\Delta $-configuration is formed. Different from
$\Lambda$-atoms~\cite{jpm,scullybook}, the energies of the
adiabatic states in this $\Delta $-atom are sensitive {\it not
only to the amplitude but also the total phase $\phi$ of the
pulses and detuning $\omega^{\prime}$} between different microwave
fields and atomic transition frequencies. The adiabatic condition
is strongly affected by the pulse phase $\phi$ when fixing other
pulse parameters. This pulse-phase-sensitive transition is due to
a broken symmetry, in which {\it one and two-photon processes can
coexist}. By adjusting the pulse phases, central times and
intensities, as well as the detunings, the populations of the
artificial atom can be adiabatically controlled. Therefore,
desired or target quantum states can be prepared using this
controllable pulse manipulation.

Finally, we emphasize the following: (i)  the adiabatic
manipulation can be completed in about $0.36\,\mu$s,  if the pulse
width is taken~\cite{yang} as, e.g., $\tau =60$ ns; (ii) this
artificial atom can be used to demonstrate three-level
masers~\cite{scullybook}; (iii) it can be regarded as a natural
candidate to realize the quantum heat engine proposed in
Ref.~\cite{scully}; (iv) a superposition of two upper energy
levels can be adiabatically prepared from the ground state, then a
quantum beat experiment should be accessible in this micro-chip
electric circuit; (v) when one of the Rabi frequencies, e.g.,
$\Omega _{01}(t)$, $\Omega _{12}(t)$, or $\Omega _{02}(t)$, and
the environmental effects are negligible, the $\Delta$-atom can be
reduced to either a $\Lambda$- $V$-, or $\Xi $-type
atom~\cite{scullybook}; (vi) the interaction between the quantized
microwave field and this $\Delta$-atom can generate quasi- and
non-classical photon states~\cite{sun}; (vii) since the total
phase of the pulses plays an important role in the state control
of $\Delta $-type atoms, this controllable pulse phase might be
used to suppress decoherence. In summary,  the artificial $\Delta
$-atoms introduced here provide many exciting future opportunities
for quantum state control.

YXL thanks A. Miranowicz for discussions. This work was supported
in part by the NSA, ARDA, under AFOSR contract
No.~F49620-02-1-0334; and also by the US NSF grant
No.~EIA-0130383. JQY was supported by the NSFC grant No.~10474013.
CPS is partially supported by the NSFC and the NFRPC with
No.~2001CB309310.


\begin{thebibliography}{99}
\bibitem{shapiro1}This phenomena has also been found in enantiomers: chiral
molecules and their mirror image, see Ref.~\cite{shapiro}.
\end{thebibliography}
\end{document}